\documentclass{article}

\usepackage{microtype}
\usepackage{graphicx}
\usepackage{subfigure}
\usepackage{booktabs} %
\usepackage{amsmath} %
\usepackage{amssymb}
\usepackage{bm}
\usepackage{float}

\usepackage{hyperref}
\usepackage{placeins} %

\usepackage[accepted]{sep-preprint}

\begin{document}

\twocolumn[
\title{Interpretable Social Network Churn Prediction\\via Ergodic Markov Chains}
\author{}
\date{\vspace{-0.2in}}
\maketitle

\begin{icmlauthorlist}
\icmlauthor{Matthew McQuistion, Microsoft Engage, mmcquistion@microsoft.com}{}
\icmlauthor{Siddhartha Sharan, Microsoft Engage, Siddhartha.Sharan@microsoft.com}{}
\icmlauthor{Afia Zaheer, Microsoft Engage, afzaheer@microsoft.com}{}
\end{icmlauthorlist}

 \icmlcorrespondingauthor{Matthew McQuistion}{mmcquistion@microsoft.com}
  \icmlcorrespondingauthor{Siddhartha Sharan}{Siddhartha.Sharan@microsoft.com}
   \icmlcorrespondingauthor{Afia Zaheer}{afzaheer@microsoft.com}

\hypersetup{
  pdftitle={Interpretable Social Network Churn Prediction via Ergodic Markov Chains},
  pdfauthor={Matthew McQuistion, Siddhartha Sharan, Afia Zaheer},
  pdfsubject={Author manuscript accepted for publication in Microsoft Journal of Applied Research, Volume 23, Article 217},
  pdfkeywords={Markov chains, interpretable models, churn prediction, social networks}
}
\begin{center}
\small Author manuscript. Accepted for publication in\\
\textit{Microsoft Journal of Applied Research}, Vol.~23, Article~217.
\par\vspace{0.5em}
\textcopyright{} 2026 Microsoft. All rights reserved.
\end{center}

\vspace{0.4in}
\begin{abstract}
Predicting social network churn before it becomes visible in aggregate metrics remains an open challenge, particularly in enterprise settings where heterogeneous usage patterns, sparse data, and deployment constraints preclude standard supervised approaches. We formalize this problem as finite-step engagement projection over contributor-role Markov chains, with a stationary-distribution interpretation for ergodic chains in the long-run limit. For each engagement feature type within a network (e.g., Q\&A, communities, articles, etc.), users are classified monthly into four behavioral states and a transition matrix is estimated from six months of history. Six-step propagation yields a label-free projection of the contributor mix; for ergodic chains, the Perron--Frobenius theorem establishes convergence under repeated propagation. Comparing projected future engagement to current engagement produces the Health Ratio, a single score requiring no trained model, no labels, and no infrastructure beyond SQL. Validated across seven rolling prediction windows on 11{,}500+ tenant-level networks, the Health Ratio achieves mean AUC 0.917 with a 156$\times$ churn-rate gradient across risk tiers, outperforming all single-signal baselines and the growth-adjusted variant previously deployed in production. An ablation study shows that six-step propagation achieves similar churn discrimination to the eigenvector-based score. Deployed on Microsoft Engage since October 2025, the framework powers customer health monitoring and has enabled proactive intervention on at-risk networks before decline becomes visible in traditional metrics.
\\

\end{abstract}
\vspace{0.4in}
]

\clearpage

\section{Introduction}
\label{sec:intro}

Social networks operate on a flywheel: users create content, others consume and react, and those reactions draw creators back. If the flywheel reverses---creators disengage, reactions thin, remaining creators feel unheard---the network can collapse rapidly. Predicting and preventing such reversals is essential to the success of any social platform, and especially so in enterprise settings, where each tenant operates in their own independent network.

\textbf{Problem. } Microsoft Engage is one such platform: each tenant hosts a self-contained social network whose engagement patterns and feature usage vary radically across organizations. Unlike consumer networks with a single global graph, network effects are bounded at the tenant level and churn is per-network rather than platform-wide. This creates a measurement problem across tens of thousands of networks that are all structurally independent: how does one assess network health in a way that is (i)~unbiased to usage style, (ii)~predictive rather than descriptive, and (iii)~actionable enough to identify what specifically must be addressed?

The most common top-down metric, Monthly Active Users (MAU), fails all three criteria. A network with 10,000 passive viewers is fundamentally different from one with 1,000 active creators, yet raw counts conflate the two. By the time systemic MAU decline becomes obvious, it is often too late to intervene. Several constraints compound the difficulty: heterogeneous dynamics across 17 feature types (\S\ref{sec:data}), no ground-truth labels at scoring cadence, sparse data in small networks, limited model-serving infrastructure, and the need for interpretable, actionable outputs that precludes black-box classifiers.

We present \emph{Stationary Engagement Projection} (SEP), a framework which addresses these constraints by combining a role-based engagement value model with a contributor-role Markov chain projection of contributor retention. For each engagement feature type (e.g.\ community posts, Q\&A, storyline, announcements), users are classified monthly into four contributor states (Creator, Replier, Reactor, inactive/Churned). A $4{\times}4$ transition matrix is estimated from six months of behavioral history, and six-step propagation of the current distribution projects the future contributor mix.

\begin{figure}[t]
\vskip 0.15in
\begin{center}
\centerline{\includegraphics[width=0.75\columnwidth]{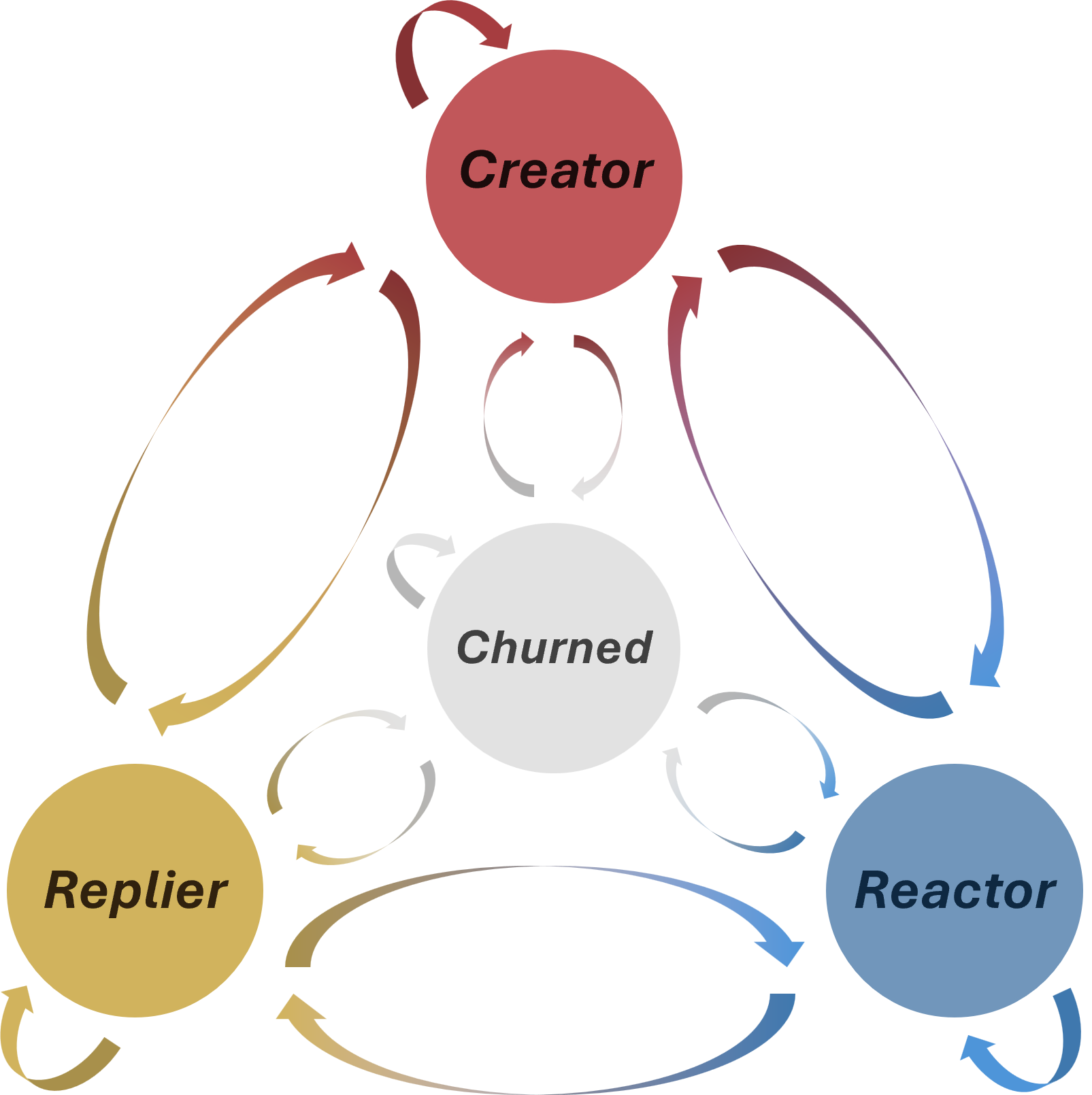}}
\caption{Diagram of contributor state Markov chain. The Churned node denotes monthly inactivity, not permanent departure.}
\label{fig:chain}
\end{center}
\vskip -0.4in
\end{figure}

\textbf{Contributions.} (1)~A formal connection between the Perron--Frobenius theorem and a productionized churn prediction model via contributor-role Markov chains; (2)~a SQL-native power iteration algorithm that achieves an AUC within 0.001 of the eigenvector-based score; (3)~multi-window temporal validation across seven rolling prediction windows (mean AUC 0.917, $\sigma = 0.006$). (4)~an ablation study validating SEP against other approaches.

\section{Related work}
\label{sec:related}

\textbf{Community survival prediction.} Hamilton et al.~\cite{hamilton2017} showed that contributor role diversity predicts Reddit community survival better than raw size---an intuition SEP operationalizes via Markov chains. Kairam et al.~\cite{kairam2012}, Backstrom et al.~\cite{backstrom2006}, and Ducheneaut et al.~\cite{ducheneaut2007} used structural graph features for community and guild longevity prediction but require labeled data. Danescu-Niculescu-Mizil et al.~\cite{danescu2013} linked linguistic change to community departure. Iriberri and Leroy~\cite{iriberri2009} proposed an online community lifecycle model; our Health Ratio quantifies lifecycle stage. Kraut and Resnick~\cite{kraut2012} identified member commitment as central to persistence.

\textbf{Markov models in retention.} Pfeifer and Carraway~\cite{pfeifer2000} formalized customer relationships as Markov chains; we adapt this to contributor roles where the ``transaction'' is a monthly engagement event. Fader et al.~\cite{fader2005} proposed the BG/NBD model for noncontractual repeat purchasing; our approach instead models monthly transitions among contributor roles. Gupta et al.~\cite{gupta2006} modeled cohort-level retention; we extend this to per-feature, per-network granularity.

\textbf{Churn prediction in practice.} Verbeke et al.~\cite{verbeke2012} found that interpretable churn models match black-box alternatives when evaluation accounts for business cost; we revisit interpretable supervised approaches---Cox proportional hazards~\cite{cox1972} and gradient boosting with SHAP attribution~\cite{lundberg2017}---in $\S$\ref{sec:discussion}. Ascarza~\cite{ascarza2018} showed that targeting high-risk customers for retention may be ineffective, motivating interpretable, per-feature risk decomposition. Wu~\cite{wu2013} established that enterprise social network structure carries information beyond individual activity metrics. To our knowledge, no scaled production system uses ergodic Markov chain projection as a forward-looking health signal. SEP has been deployed on Microsoft Engage since October 2025.

\section{Theoretical framework}
\label{sec:theory}

\subsection{State space and transition model}
\label{sec:state_space}

Let $\mathcal{S} = \{C, R, X, L\}$ denote the four contributor states: Creator ($C$), Replier ($R$), Reactor ($X$), and inactive/Churned ($L$). For a given network and feature type $f$, each user $u$ who was active in feature $f$ during any month of the six-month lookback window is assigned to exactly one state per month based on their highest-engagement action:

\begin{table}[h]
\begin{center}
\begin{small}
\begin{tabular}{ll}
\toprule
\textbf{State} & \textbf{Definition} \\
\midrule
Creator ($C$) & Posted $\geq 1$ thread starter that month \\
Replier ($R$) & Replied to a thread; did not start one \\
Reactor ($X$) & Reacted to $\geq 1$ thread or reply \\
Inactive/Churned ($L$) & In lookback; inactive this month \\
\bottomrule
\end{tabular}
\end{small}
\end{center}
\vskip -0.15in
\end{table}

The assignment is deterministic and mutually exclusive with a strict hierarchy: $C > R > X > L$. A user who both created and replied is classified as Creator; a user active in the lookback but absent this month is inactive/Churned. This monthly state permits subsequent reactivation and is distinct from the network-level churn outcome defined in Section~\ref{sec:data}.

For each (network, feature) pair, we estimate a $4{\times}4$ row-stochastic transition matrix $\mathbf{P}_f \in \mathbb{R}^{4 \times 4}$ from six steps of observed month-over-month transitions:
\begin{equation}
\label{eq:transition}
P_f(i,j) = \frac{|\{u : s_u(t) = i,\; s_u(t{+}1) = j\}|}{\sum_{j' \in \mathcal{S}} |\{u : s_u(t) = i,\; s_u(t{+}1) = j'\}|},
\end{equation}
pooled over all months $t$ in the lookback window. For states with observed outgoing transitions, probabilities are estimated using the count ratio above. When no outgoing transitions are observed, we set the self-transition probability to one and all other entries in that row to zero. Each row therefore sums to one: $\sum_j P_f(i,j) = 1$ for all $i$.

\textbf{Why monthly aggregation.} Monthly windows balance signal stability against responsiveness: weekly windows produce unstable state assignments in small networks where users post in only a handful of weeks per quarter, while quarterly aggregation forfeits the leading-indicator behavior described in \S\ref{sec:discussion}. Monthly cadence also aligns with the natural rhythm of enterprise engagement (leader posts, newsletters, town halls) and Engage's existing metric review schedule. 

\textbf{Why per-feature matrices.} Transition dynamics are structurally heterogeneous across feature types. Community posts exhibit balanced role retention (Creator self-transition $p_{CC} = 0.32$, spectral gap $\gamma = 0.71$), while AMA live events are episodic with 86\% inactive/Churned between events ($\gamma = 0.36$). Collapsing these into a single global matrix would average over qualitatively different dynamics; per-feature estimation preserves each feature's independent structural trajectory and allows for additional analytical segmentation.

\subsection{The Perron--Frobenius theorem and stationarity}
\label{sec:perron_frobenius}

The theoretical foundation of our approach rests on the Perron--Frobenius theorem for non-negative matrices~\cite{seneta2006,horn2013}; we follow the spectral gap framework of Levin and Peres~\cite{levin2017} for convergence analysis.
\\
\\
\textbf{Theorem (Perron--Frobenius).} Let $\mathbf{P}$ be an irreducible, aperiodic ($\therefore$ ergodic) row-stochastic matrix. Then $\mathbf{P}$ has a unique stationary distribution $\bm{\pi}^*$ satisfying $\bm{\pi}^* = \bm{\pi}^* \mathbf{P}$, with all entries strictly positive: $\pi^*_i > 0$ for all $i \in \mathcal{S}$. Moreover, for any initial distribution $\bm{\pi}_0$:
\begin{equation}
\label{eq:convergence}
\lim_{k \to \infty} \bm{\pi}_0 \mathbf{P}^k = \bm{\pi}^*,
\end{equation}
with geometric convergence governed by the spectral gap.

\textbf{Spectral gap and convergence rate.} For ergodic chains, the magnitudes of the non-unit eigenvalues characterize asymptotic convergence~\cite{kemeny1976}. The spectral gap is one minus the largest of these magnitudes: $\gamma = 1 - |\lambda_2|$. Finite-step approximation error also depends on the transition matrix and the initial distribution; the spectral gap alone does not establish that six steps are sufficient.

\begin{figure*}[t]
\vskip 0.1in
\begin{center}
\centerline{\includegraphics[width=\textwidth]{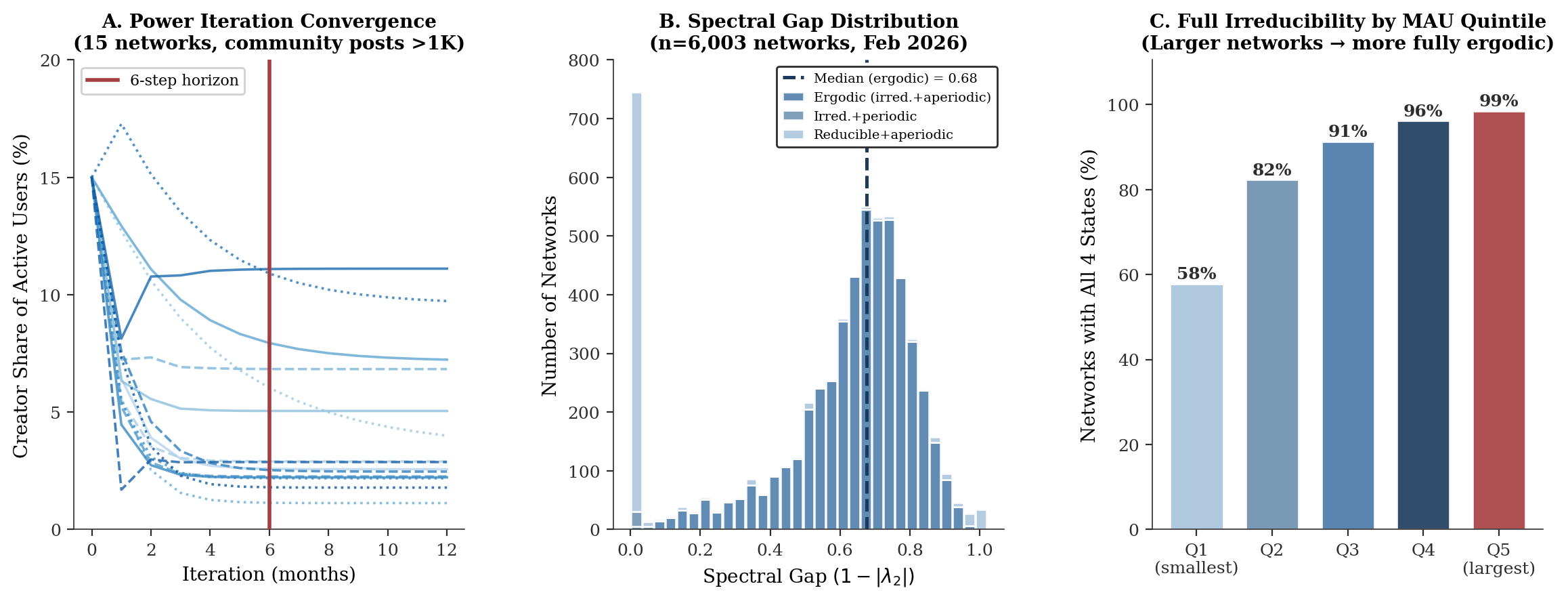}}
\caption{Markov chain properties. (A)~Creator-share trajectories for 15 representative networks under pure power iteration. (B)~Spectral gap distribution ($n{=}6{,}003$); median $\gamma{=}0.68$. (C)~Full four-state irreducibility rate by MAU quintile: 99\% for the largest networks vs.\ 58\% for the smallest. Panel C reports irreducibility, which alone does not establish aperiodicity.}
\label{fig:markov}
\end{center}
\vskip -0.2in
\end{figure*}

\textbf{Empirical chain properties.} For networks with sufficient transition data in the dominant feature type (\texttt{community\_posts\_1k\_plus}, $n{=}6{,}003$, Feb.\ 2026), 85\% have fully irreducible matrices and the median spectral gap is $\gamma = 0.68$. Separately, the ablation comparison shows similar churn discrimination for six-step projection and the eigenvector-based score ($\Delta\text{AUC} = 0.001$). This predictive agreement does not by itself establish convergence of the projected distributions. Figure~\ref{fig:markov} illustrates Creator-share trajectories, the spectral gap distribution, and the relationship between size and irreducibility.

\textbf{Handling non-ergodic chains.} SEP propagates the observed contributor distribution through six transition steps. This calculation remains well-defined for non-ergodic chains and preserves dependence on the initial distribution, but its output need not be stationary. The self-loop convention for unobserved rows makes those states absorbing in the estimated matrix; it does not establish permanent inactivity in the underlying population.

\subsection{Engagement value model}
\label{sec:value_model}

Not all contributor states generate equal engagement value, and they do so through different mechanisms. We decompose the feature-level score into an \emph{activity matrix}, a \emph{value vector}, and the state distribution.

\textbf{Activity matrix.} For each feature type $f$, we define $\mathbf{A}_f \in \mathbb{R}^{3 \times 3}$ as the activity matrix whose rows correspond to contributor states $\{C, R, X\}$ and columns to contribution event classes (thread starts, replies, reactions). Each entry $A_f(s, c)$ records the expected number of contribution events of class $c$ per user in state $s$ per month:
\begin{equation}
\label{eq:activity_matrix}
\mathbf{A}_f = \begin{pmatrix}
\mathbb{E}[\text{threads}|C] & \mathbb{E}[\text{replies}|C] & \mathbb{E}[\text{reacts}|C] \\
0 & \mathbb{E}[\text{replies}|R] & \mathbb{E}[\text{reacts}|R] \\
0 & 0 & \mathbb{E}[\text{reacts}|X]
\end{pmatrix}\!
\end{equation}
The upper-triangular sparsity follows from the state hierarchy ($C > R > X$): Reactors produce only reactions, Repliers produce replies and reactions, and Creators participate across all event classes. Inactive/Churned users are omitted as they generate no value in the period.

\textbf{Value vector.} Let $\mathbf{v}_f = (v_{\text{thread}}, v_{\text{reply}}, v_{\text{react}})^\top \in \mathbb{R}^3$ denote the incremental value from one contribution event of each class, estimated from six months of network-level reaction ratios (downstream reactions per thread, per reply, per direct reaction).

\textbf{Feature score.} The short-term feature score is then:
\begin{equation}
\label{eq:value}
s_f = \bm{\pi}_{f}^{\,\text{act}} \, \mathbf{A}_f \, \mathbf{v}_f,
\end{equation}
where $\bm{\pi}_{f}^{\,\text{act}} = (\pi_C, \pi_R, \pi_X) \in \mathbb{R}^3$ denotes the active-state entries of the full distribution $\bm{\pi}_f \in \mathbb{R}^4$. This factorization separates who is present ($\bm{\pi}_f$), what they do ($\mathbf{A}_f$), and what it is worth ($\mathbf{v}_f$). The log transform applied at the aggregate level (Eqs.~\ref{eq:network_score}--\ref{eq:health_ratio}) compresses magnitude differences across networks.

Empirically, the value model contributes to discriminative performance: replacing the learned $\mathbf{A}_f$ and $\mathbf{v}_f$ with uniform weights (equal value per state) degrades AUC by 0.007 across the seven temporal windows and by 0.004 on the ablation cohort (Table~\ref{tab:comparison}).

\subsection{SEP computation}
\label{sec:score}
\textbf{Six-step projected score.} Given $s_f$ from Eq.~\ref{eq:value}, we project the full four-state distribution forward six steps:
\begin{equation}
\label{eq:power_iter}
\bm{\pi}_f^{(6)} = \bm{\pi}_f \, \mathbf{P}_f^6,
\end{equation}
and compute the projected engagement value from the active entries:
\begin{equation}
\label{eq:lt}
s_f^* = \bm{\pi}_{f}^{(6),\text{act}} \, \mathbf{A}_f \, \mathbf{v}_f.
\end{equation}

\textbf{MAU-weighted aggregation.} Feature-level scores are combined using feature MAU share as weights:
\begin{equation}
\label{eq:agg}
\sigma = \sum_f w_f \cdot s_f, \quad \sigma^* = \sum_f w_f \cdot s_f^*,
\end{equation}
where $w_f = \mathrm{MAU}_f / \sum_{f'} \mathrm{MAU}_{f'}$.

\textbf{Network Score and Health Ratio.} The aggregated scores yield two complementary metrics. The \emph{Network Score} $\mathcal{N}$ combines current and projected engagement into a single composite:
\begin{equation}
\label{eq:network_score}
\mathcal{N} = \ln(1 + \sigma + \sigma^*).
\end{equation}
The \emph{Health Ratio} $\mathcal{R}$ isolates the structural trajectory by comparing projected to current engagement:
\begin{equation}
\label{eq:health_ratio}
\mathcal{R} = \frac{\ln(1 + \sigma^*)}{\ln(1 + \sigma)}.
\end{equation}
Networks with $\sigma = 0$ are assigned $\mathcal{R} = 0$ (Severe by convention, ${<}0.1\%$ of cases). $\mathcal{R} > 1$ indicates structural improvement---above-average transitions into higher-value states outweigh the current distribution---and is uncommon; generally indicating that the network's latest-month contributor mix is an outlier, thus the projection reflects mean-regression. Although $\mathcal{N}$ is used operationally, this paper evaluates $\mathcal{R}$ as the primary churn predictor.

\section{Algorithm and implementation}
\label{sec:algorithm}

We summarize the end-to-end SEP pipeline in Algorithm~\ref{alg:network_score}: per-(user, feature, month) state assignment feeds transition estimation, six propagation steps project the contributor distribution, and the per-feature scores are MAU-weighted into the network-level Health Ratio.

\begin{algorithm}[hbt!]
   \caption{Stationary Engagement Projection}
   \label{alg:network_score}
\begin{algorithmic}
   \STATE {\bfseries Input:} Six-month event history per (network, $f$, $u$, $t$)
   \STATE {\bfseries Output:} Health Ratio $\mathcal{R}$ and risk tier per network
   \STATE
   \STATE \textbf{1.} Assign contributor state $s_u^{(f,t)} \in \{C,R,X,L\}$ for each user $u$, feature $f$, month $t$
   \STATE \textbf{2.} Estimate $4{\times}4$ transition matrix $\mathbf{P}_f$ via Eq.~\ref{eq:transition}, with self-loops for unobserved rows ($\S$\ref{sec:state_space})
   \STATE \textbf{3.} Estimate $\mathbf{A}_f$ and $\mathbf{v}_f$ from engagement data ($\S$\ref{sec:value_model})
   \STATE \textbf{4.} Compute current distribution $\bm{\pi}_f$ from latest month
   \STATE \textbf{5.} Compute $s_f = \bm{\pi}_{f}^{\,\text{act}} \, \mathbf{A}_f \, \mathbf{v}_f$
   \FOR{$k=1$ {\bfseries to} $6$}
   \STATE $\bm{\pi}_f \leftarrow \bm{\pi}_f \, \mathbf{P}_f$ \quad (matrix--vector multiply)
   \ENDFOR
   \STATE \textbf{6.} Compute $s_f^* = \bm{\pi}_{f}^{(6),\text{act}} \, \mathbf{A}_f \, \mathbf{v}_f$
   \STATE \textbf{7.} Aggregate: $\sigma = \sum_f w_f \cdot s_f$, $\sigma^* = \sum_f w_f \cdot s_f^*$
   \STATE \textbf{8.} $\mathcal{R} \leftarrow \ln(1{+}\sigma^*) \,/\, \ln(1{+}\sigma)$
   \STATE \textbf{9.} Assign risk tier from $\mathcal{R}$ (Table~\ref{tab:tiers})
\end{algorithmic}
\end{algorithm}

\textbf{SQL-native power iteration.} Each power iteration step (the inner loop of Algorithm~\ref{alg:network_score}) is a single aggregated join expressible as a materialized common table expression (CTE). For state $j$ at step $k{+}1$:
\begin{equation}
\label{eq:sql_step}
\pi_j^{(k+1)} = \sum_{i \in \mathcal{S}} \pi_i^{(k)} \cdot P_f(i,j).
\end{equation}
This is a dot product between the current distribution vector and column $j$ of $\mathbf{P}_f$, implemented as a \texttt{SUM} over a join on state index. Six sequential CTEs yield the full projection---no matrix factorization, no iterative solver, and no external runtime~\cite{golub2013}, integrating directly into existing analytical pipelines.

\textbf{Risk tier definition.} Tier boundaries were selected from natural breakpoints in observed churn rates in the Sep 2025 $\rightarrow$ Feb 2026 cohort (Table~\ref{tab:tiers}). These outcome-informed boundaries assign descriptive risk labels but do not alter the continuous Health Ratio. The same boundaries are used across the seven rolling evaluation windows ($\S$\ref{sec:temporal}). Because these windows overlap, they should not be interpreted as independent validations of the tier boundaries.
\begin{table}[!htbp]
\begin{center}
\begin{small}
\begin{sc}
\begin{tabular}{lc}
\toprule
Health Ratio & Risk Tier \\
\midrule
$\geq 0.55$ & No Risk \\
$0.48$--$0.55$ & Low \\
$0.39$--$0.48$ & Moderate \\
$0.28$--$0.39$ & High \\
$< 0.28$ & Severe \\
\bottomrule
\end{tabular}
\end{sc}
\end{small}
\end{center}
\caption{Risk tier definition. Boundaries derived from natural breakpoints in the churn-rate gradient (Sep 2025 $\rightarrow$ Feb 2026 cohort, $n{=}11{,}941$, ${\geq}100$ MAU); observed 5-month churn rates per tier are reported in Figure~\ref{fig:main}B.}
\label{tab:tiers}
\vskip -0.2in
\end{table}

\textbf{Computational cost.} The full pipeline of state classification, transition estimation, six power iteration steps, scoring, and aggregation executes for all global networks in under 30 minutes on a Snowflake SQL warehouse, orchestrated via Apache Airflow DAGs.

\section{Data}
\label{sec:data}

\textbf{Source and scope.} All event data (thread starts, replies, reactions, views) from the production telemetry pipeline of Microsoft Engage, aggregated at the (network, feature\_type, user, month) level before entering the scoring pipeline. 17 feature types are supported, spanning community posts across three size tiers, Q\&A, storyline posts, campaigns, announcements, and others.

\textbf{Scale.} 612,495 network-month observations across 49,695 networks over 22 months (May 2024--February 2026). Network MAU ranges from under 10 to over 440,000. For temporal validation, seven rolling six-month prediction windows span March through September 2025, yielding 11,487 to 11,941 networks per window at ${\geq}100$ MAU.

\textbf{Score timing.} Scores are labeled by their scoring month but use only activity from months preceding the score month; for example, an October score contains no October activity. We distinguish this score-month convention from the five-month outcome horizon.

\textbf{Churn definition.} A network is considered churned if it drops below 100 MAU or loses ${\geq}90\%$ of MAU over five months. This outcome label is not used to compute the continuous Health Ratio. It is used for evaluation and for selecting the descriptive risk-tier boundaries. No individual user data is retained: all analysis runs on aggregate state-count tables.

\section{Results}
\label{sec:results}

\subsection{Multi-window temporal validation}
\label{sec:temporal}

To assess temporal stability, we evaluate across seven rolling six-month prediction windows (March 2025--Sep 2025 start months, 11{,}487--11{,}941 networks per window at ${\geq}100$ MAU). For each window, we recompute the no-growth Health Ratio from raw transition matrices. Mean ROC AUC = 0.917 ($\sigma = 0.006$, range $[0.910, 0.928]$); the tight variance indicates stable discrimination across these rolling windows without recalibrating the continuous score. The span includes summer months (Jun--Aug starts) when vacation seasonality shifts enterprise engagement, and churn base rates remain stable across windows (${\sim}4.8\%$). For comparison, the growth-adjusted production variant achieves mean AUC 0.824 ($\sigma = 0.018$) over the same windows. The $\Delta\text{AUC} = -0.093$ penalty that confirms the single-window ablation finding ($\S$\ref{sec:ablation}) holds across all temporal periods.

\subsection{Churn rate by risk tier}
\label{sec:tiers}

\begin{figure*}[t]
\begin{center}
\centerline{\includegraphics[width=\textwidth]{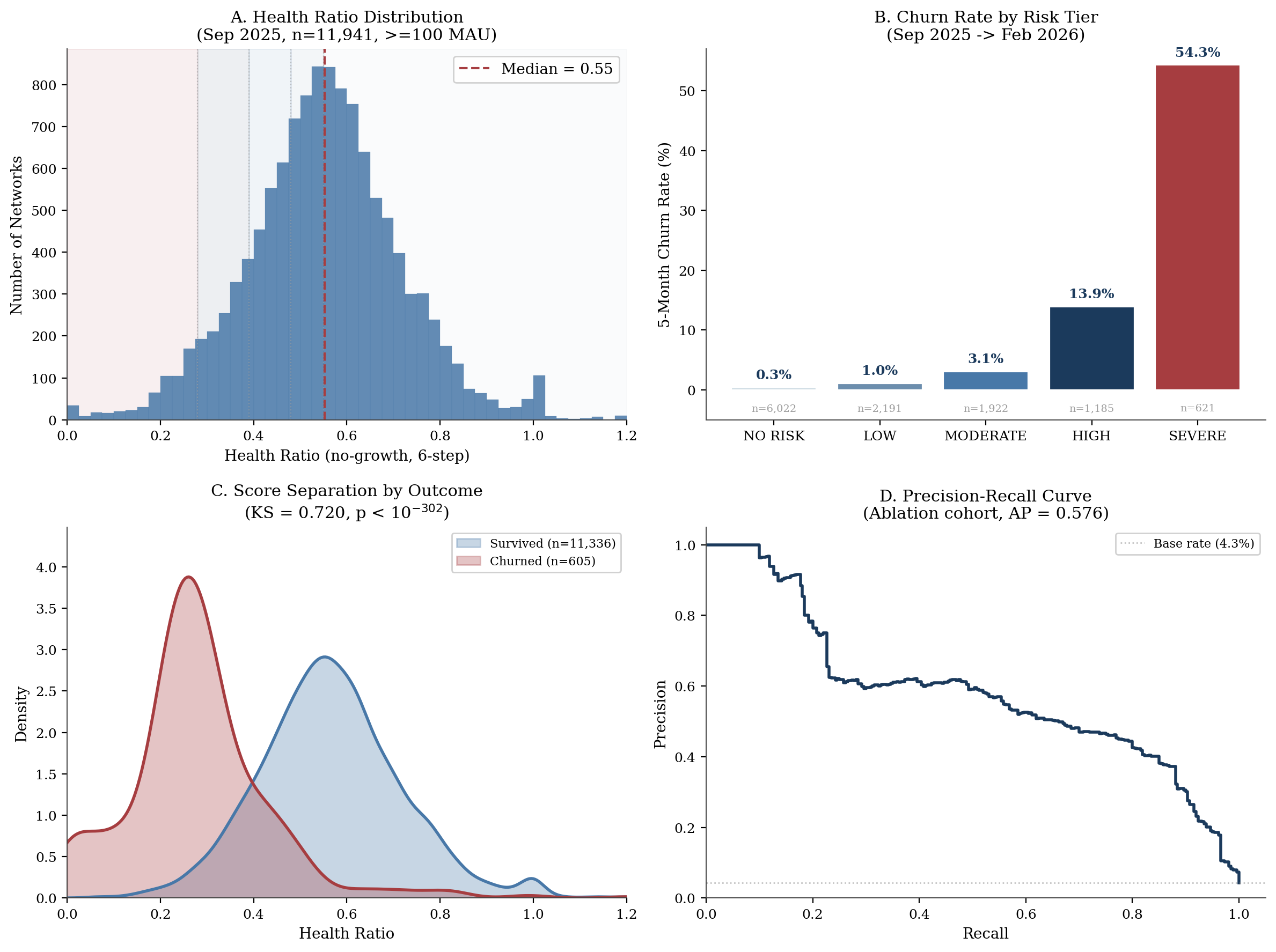}}
\caption{Main results. (A)~Health Ratio distribution with risk tier boundaries ($n{=}11{,}941$, ${\geq}100$ MAU); median $= 0.55$. (B)~Observed five-month churn rate by risk tier: 156$\times$ gradient from No Risk (0.3\%) to Severe (54.3\%). (C)~Score separation by outcome; KS $= 0.720$, $p < 10^{-302}$. (D)~Precision-recall (ablation cohort, $n{=}6{,}092$); AP $= 0.576$ ($13{\times}$ base rate).}
\label{fig:main}
\end{center}
\vskip -0.25in
\end{figure*}

Five-month outcomes for the Sep 2025 $\rightarrow$ Feb 2026 cohort ($n = 11{,}941$ networks, ${\geq}100$ MAU, 605 churned, 5.1\% base rate) are shown in Figure~\ref{fig:main}. The churn-rate gradient is strictly monotone: Severe networks churn at 156$\times$ the rate of No Risk networks (54.3\% vs.\ 0.3\%). Among networks with complete transition data for the dominant feature type ($n{\approx}6{,}000$; see $\S$\ref{sec:ablation}), 87\% of churning networks fall in the lowest Health Ratio decile.

\subsection{Risk stratification and operational lift}
\label{sec:calibration}

The Health Ratio separates networks into groups with substantially different observed churn rates. Across the five risk tiers in Figure~\ref{fig:main}B, observed five-month churn increases monotonically from 0.3\% to 54.3\%. These rates describe outcomes within the selected tiers; they do not establish that the Health Ratio is a calibrated churn probability. Tier boundaries were selected using the cohort described in Section~\ref{sec:algorithm}. The cumulative lift curve shows that flagging the top 10\% of at-risk networks captures 52\% of all eventual churners (${\sim}5{\times}$ lift over random), providing a practical operating point for resource-constrained intervention programs.

The Health Ratio also produces well-separated score distributions for churners vs.\ survivors (Figure~\ref{fig:main}C): KS $= 0.720$, $p < 10^{-302}$, with the median churner Health Ratio (0.27) falling below the 5th percentile of survivor scores (0.33). Under the 4.3\% ablation-cohort base rate, precision-recall analysis is more informative than ROC alone~\cite{davis2006}; the precision-recall curve (Figure~\ref{fig:main}D) achieves average precision 0.576 ($13{\times}$ enrichment over random).

\subsection{Network size sensitivity}
\label{sec:size}

AUC improves with network size (Table~\ref{tab:size}), driven by two reinforcing factors: larger networks produce more reliable transition matrix estimates (more observed transitions per cell), and they exhibit higher irreducibility rates (all four states are represented in the transition data). At ${\geq}1{,}000$ MAU, 83\% of networks have fully irreducible matrices vs.\ 45\% across all networks. Degradation is graceful: even at ${\geq}100$ MAU (the primary evaluation threshold), mean AUC remains high at 0.917.
\begin{table}[!htbp]
\begin{center}
\begin{small}
\begin{sc}
\begin{tabular}{lccr}
\toprule
Min MAU & Mean AUC & Churn Rate & $n$/window \\
\midrule
$\geq$ 100   & 0.917 & 4.8\% & $\sim$11,800 \\
$\geq$ 500   & 0.951 & 1.5\% & $\sim$4,900 \\
$\geq$ 1,000 & 0.971 & 1.0\% & $\sim$3,200 \\
\bottomrule
\end{tabular}
\end{sc}
\end{small}
\end{center}
\caption{AUC by minimum MAU threshold (mean across 7 windows, no-growth Health Ratio). Larger networks yield more reliable transition estimates and higher irreducibility rates.}
\label{tab:size}
\vskip -0.2in
\end{table}

\subsection{Ablation and baseline comparisons}
\label{sec:ablation}

\begin{figure}[t]
\begin{center}
\centerline{\includegraphics[width=\columnwidth]{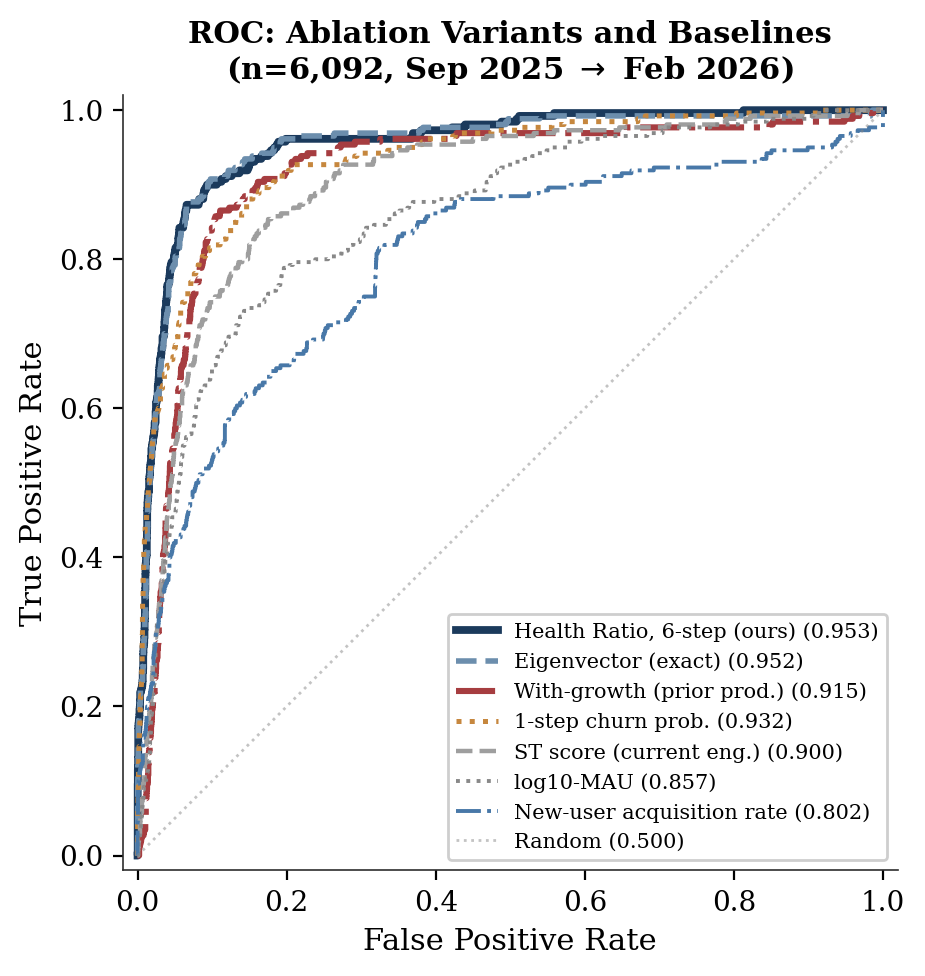}}
\caption{ROC curves for ablation variants and baselines (ablation cohort, $n{=}6{,}092$, Sep 2025 $\rightarrow$ Feb 2026). The legend's ``Eigenvector (exact)'' denotes the eigenvector-based score; ``1-step churn prob.'' denotes the projected inactive/Churned share, not a calibrated network-churn probability.}
\label{fig:roc}
\end{center}
\vskip -0.2in
\end{figure}

We compare the Health Ratio against ablation variants and non-Markov baselines (Figure~\ref{fig:roc}, Table~\ref{tab:comparison}). Computing exact eigenvectors and growth-adjusted scores requires raw transition matrices, restricting the ablation to $n{=}6{,}092$ networks with sufficient data in the dominant feature type (of 11,941 at ${\geq}100$ MAU). The excluded networks are predominantly small (${<}200$ MAU). The ablation cohort's AUC (0.953) is consistent with the seven-window average for similarly sized networks (Table~\ref{tab:size}).

\begin{table}[!htbp]
\begin{center}
\begin{small}
\begin{sc}
\begin{tabular}{lc}
\toprule
Method & AUC \\
\midrule
\multicolumn{2}{l}{\textit{SEP variants}} \\
\quad Health Ratio, 6-step \emph{(ours)} & \textbf{0.953} \\
\quad Eigenvector-based score & 0.952 \\
\quad Uniform value weights & 0.949 \\
\quad With-growth \emph{(prior prod.)} & 0.915 \\
\quad \shortstack[l]{1-step projected\\inactive/Churned share} & 0.932 \\
\midrule
\multicolumn{2}{l}{\textit{Non-Markov baselines}} \\
\quad ST score (current engagement) & 0.900 \\
\quad $\log_{10}$-MAU & 0.857 \\
\quad New-user acquisition rate & 0.802 \\
\midrule
\multicolumn{2}{l}{\textit{Supervised (5-fold CV)}} \\
\quad Logistic reg.\ (4 features) & 0.970 \\
\bottomrule
\end{tabular}
\end{sc}
\end{small}
\end{center}
\caption{Ablation and baseline comparison (Sep 2025 $\rightarrow$ Feb 2026, $n{=}6{,}092$ networks with complete transition data). AUCs are consistent with the seven-window temporal averages for similarly sized networks (cf.\ Table~\ref{tab:size}).}
\label{tab:comparison}
\vskip -0.1in
\end{table}

\textbf{Ablation.} The six-step and eigenvector-based scores achieve AUCs of 0.953 and 0.952, respectively, indicating similar discrimination in this cohort. This comparison does not measure stationary-distribution approximation error. The growth-adjusted variant previously deployed in production trails by 0.038 AUC on this cohort, and by 0.093 AUC across the full seven-window temporal validation ($\S$\ref{sec:temporal}), because compounding inflation disproportionately raises long-term scores for structurally weak but high-growth networks.

\textbf{Baselines.} The strongest unsupervised baseline is the one-step projected inactive/Churned share ($(\bm{\pi}_f \mathbf{P}_f)_L$, MAU-weighted; AUC 0.932), confirming that the transition matrix itself carries substantial signal. The Health Ratio's additional 0.021 AUC derives from the six-step projection, which distinguishes transient dips from structural trajectory toward the inactive/Churned state. Current engagement (ST score, 0.900) and network size ($\log$-MAU, 0.857) are both predictive but capture only present-tense information; the Health Ratio adds a forward-looking structural component.

\textbf{Supervised upper bound.} A logistic regression trained on four features (Health Ratio, ST score, $\log$-MAU, new-user rate) achieves 0.970 AUC under 5-fold cross-validation (+0.017 over $\mathcal{R}$ alone), with the Health Ratio coefficient $2.5{\times}$ larger than any other feature. This establishes that supervised methods can extract meaningful additional value by combining SEP outputs with complementary signals.

\section{Discussion}
\label{sec:discussion}

\textbf{Power iteration vs.\ eigendecomposition.} Finite-step propagation is directly expressible in SQL and retains dependence on the observed initial distribution. For non-ergodic chains, a unique stationary distribution or convergence to a stationary limit is not guaranteed. SEP therefore uses the six-step projection rather than requiring a stationary solution. In the evaluated ablation cohort, its discrimination is similar to that of the eigenvector-based score.

\textbf{Leading indicator behavior.} The Health Ratio is not merely a point-in-time classifier: longitudinal analysis over the 22-month observation window (\S\ref{sec:data}) shows that eventual churners exhibit significantly lower Health Ratios eight months before churning (mean $\Delta \mathcal{R} = 0.111$ at $t{-}8$), widening monotonically to $\Delta \mathcal{R} = 0.283$ at the prediction point. Autocorrelation decays from $r = 0.695$ (lag~1) to $r = 0.433$ (lag~3), confirming a persistent signal that still responds to monthly behavioral changes.

\textbf{Strengths.} The framework is fully interpretable: every component maps to a specific feature type and contributor role, enabling targeted intervention (e.g., ``Community Posts creators are churning faster than repliers''), complementing optimization approaches that model intervention effects on social activity~\cite{farajtabar2014}. It is privacy-preserving (aggregate-only SQL, no individual records retained); extensible (new feature types require only a post-type flag, no retraining); and computationally efficient. Health Ratio's weak correlation with $\log$-MAU ($r = 0.06$) confirms it captures structural information beyond size.

\textbf{Relationship to discriminative interpretable models.} SEP complements rather than competes with discriminative interpretable approaches such as gradient boosting with SHAP attribution~\cite{lundberg2017} or Cox proportional hazards~\cite{cox1972}. Where these methods fit a labeled distribution and attribute predictions post-hoc, SEP is a mechanistic, generative model whose interpretability is constructive in the sense of Rudin~\cite{rudin2019}: each modeled quantity (transition probabilities, projected contributor mix, activity matrix, Health Ratio) corresponds directly to a measurable property of the system rather than to an attribution computed against a fitted predictor. However, these structural advantages come at one structural cost: arbitrary nonlinear interactions among raw features cannot be learned (\S\ref{sec:ablation}, supervised upper bound). The two compose naturally: SEP's per-feature Health Ratios, projected distributions, and spectral gaps can serve as engineered features for downstream supervised models, yielding attributions in terms of structural dynamics rather than raw event counts.

\textbf{Limitations.} The model assumes transition dynamics are approximately stationary over the six-month lookback window; rapidly shifting engagement patterns (e.g., a viral event that permanently changes the contributor mix) may violate this assumption, though the monthly re-scoring cadence bounds the impact. Networks below 100 MAU produce sparse transition matrices and noisier estimates. The score captures engagement structure, not content quality. A network with high-quality content but low structural engagement may be underrated. Finally, per-feature transition matrices are estimated independently and aggregated via MAU weighting; cross-feature dependencies are captured implicitly rather than jointly modeled, motivating future tensor-based extensions.

\section{Conclusion}
\label{sec:conclusion}

We have shown that contributor-role Markov chain projection provides a principled, label-free, and SQL-deployable framework for social network churn prediction. Across seven rolling temporal windows, the Health Ratio achieves mean AUC 0.917 ($\sigma = 0.006$) without recalibrating the continuous score. Six-step propagation achieves similar discrimination to the eigenvector-based score and outperforms the growth-adjusted production variant ($\Delta\text{AUC} = 0.093$) and all single-signal baselines. The Perron--Frobenius theorem establishes convergence for ergodic chains, while finite-step propagation remains applicable to non-ergodic chains without requiring a stationary limit. The factored value model provides the interpretability that black-box alternatives lack.

Deployed on Microsoft Engage since October 2025, SEP powers an internal dashboard used by product teams to monitor tenant health, identify underperformers, and prioritize targeted interventions. Surfacing structured risk signals has enabled proactive outreach before decline becomes visible in aggregate metrics. Work is underway to combine SEP outputs with LLM-based analysis to generate automated, tenant-specific recommendations.

The framework generalizes across the Microsoft 365 collaboration surface: channel and chat retention in Teams, contributor flow on SharePoint and Loop, etc. SEP only requires role-based engagement and periodic observability. These conditions are common in enterprise settings, where organizational structure and workflow conventions naturally partition users into hierarchical contributor states. By grounding network health in a mechanistic framework rather than platform-specific heuristics, SEP offers a reusable, actionable, and interpretable foundation for churn prediction across multi-user collaborative systems.
\FloatBarrier
\begingroup
\small
\bibliographystyle{ieeetr}
\bibliography{msjar_refs}
\endgroup

\end{document}